\documentclass[twocolumn, 3p]{elsarticle}
\usepackage{lineno,hyperref}
\usepackage{amsmath}
\usepackage[figurename=Fig., tablename=Table]{caption}

\modulolinenumbers[1]
\journal{\rm{Nucl. Instr. Meth. A}}
\begin{document}
\begin{frontmatter}

%
%
\title{\flushleft{Development of an alpha-particle imaging detector based on a low 
radioactivity
micro-time-projection chamber}}

\author{
H. Ito$^{a*}$,            
T. Hashimoto$^{a}$, 
K. Miuchi$^{a}$,       
K. Kobayashi$^{b, c}$,
Y. Takeuchi$^{a, c}$,       
K. D. Nakamura$^{a}$,
T. Ikeda$^{a}$, 		
and H. Ishiura$^{a}$ 
}
\address{$^{a}$Kobe University, Kobe, Hyogo 657-8501, Japan.}
\address{$^{b}$Institute for Cosmic Ray Research (ICRR), the University of Tokyo, Kashiwa, Chiba 277-8582 Japan.}
\address{$^{c}$Kavli Institute for the Physics and Mathematics of the Universe (WPI), The University of Tokyo Institutes for Advanced Study, University of Tokyo, Kashiwa, Chiba 277-8583, Japan.}
\cortext[cor1]{Corresponding author. E-mail address: ito.hiroshi@crystal.kobe-u.ac.jp (H. Ito).}

%
%
\begin{abstract}
An important issue for rare-event-search experiments, such as the search for dark matter or neutrinoless double beta decay, is to reduce radioactivity of the detector materials and the experimental environment.
The selection of materials with low radioactive impurities, such as isotopes of the uranium and thorium chains, requires a precise measurement of surface and bulk radioactivity.
Focused on the first one, an alpha-particle detector has been developed based on a gaseous micro-time-projection chamber.
A low-$\alpha$ $\mu$-PIC with reduced alpha-emission background was installed in the detector.
The detector offers the advantage of position sensitivity, which allows the alpha-particle contamination  of the sample to be imaged and the background to be measured at the same time.
The detector performance was measured by using an alpha-particle source.
The measurement with a sample was also demonstrated and the sensitivity is discussed.
\end{abstract}
 
%
%
\end{frontmatter}

%
%

\section{Introduction}

Approximately 27\% of the universe is dominated by non-baryonic matter, called dark matter.
Although many experimental groups have been 
searching for dark matter, arguably a direct detection has not been observed.
Typical experiments that search for dark matter are performed by using massive, low-background detectors.
Although the DAMA group has observed the presumed annual modulation of dark matter particles in the galactic halo with a significance of ${9.3\sigma}$ \cite{DAMA2018}, other groups such as XENON1T \cite{XENO1T_2017} and LUX \cite{LUX2017} 
were unable to confirm these
results.
Meanwhile, a direction-sensitive method has been focused because of an expected clear anisotropic signal due to the motion of the solar system in the galaxy \cite{WIMP-wind}.
The NEWAGE group precedes a three-dimensionally sensitive dark matter search with a micro-time-projection chamber
(micro-TPC), being the main background surface alpha particles
from $^{238}$U and $^{232}$Th in the detector materials or in the $\mu$-PIC \cite{NEWAGE2015}.

Neutrinoless double beta $(0\nu\beta\beta)$ decay is a lepton-number-violating process, which suggests the neutrino as a Majorana particle (i.e. it is its own antiparticle).
Experiments like GERDA \cite{GERDA2017} and KamLAND-Zen \cite{KamLAND2016} have been able to set a lower limit on the half-life over
${10^{25}\;{\rm yr}}$ and ${10^{26}\;{\rm yr}}$
at 90\%CL by using $^{76}$Ge and $^{136}$Xe, respectively, but no positive signal of the $0\nu\beta\beta$ process has
been observed yet.
Conversely, a tracking system for two electrons provides strong evidence of the $0\nu\beta\beta$ decay process.
The $0\nu\beta\beta$ background has been well investigated as radioactive impurities such as $^{238}$U and $^{232}$Th decay-chain isotopes, $^{40}$K, $^{60}$Co, $^{137}$Cs including in the detector material, which emit $\gamma$ with around MeV \cite{EXO2017, MAJO2016}.
The NEMO3 group 
set lower limits at
${T_{1/2}(0\nu\beta\beta)>2.5\times10^{23}\;\rm{yr}}$ (90\%CL) for $^{82}$Se \cite{NEMO3.2018}, and ${T_{1/2}(0\nu\beta\beta)>(1.1-3.2)\times10^{21}\;\rm{yr}}$ (90\%CL) for $^{150}$Nd \cite{NOME3.2017}.
For this experiment background is dominated by the $^{208}$Tl and $^{214}$Bi contamination present in the double beta emitter source foils.
The SuperNEMO group has developed the BiPo-3 detector to measure the radioactive 
impurities in these foils with a sensitivity
less than  ${2\;\rm{\mu Bq/kg}}$ (90\%CL) for $^{208}$Tl and ${140\;\rm{\mu Bq/kg}}$ (90\%CL) for $^{214}$Bi \cite{SuperNOME.BiPo.2017}.
Therefore, the background of $0\nu\beta\beta$ decay is not only a contamination by the end point of continuous energy in an ordinary $2\nu\beta\beta$ decay process, but also the radiative impurities such as $^{238}$U and $^{232}$Th in the detector.

To estimate the radioactive impurities in the detector materials, the XMASS group measured $^{210}$Pb and $^{210}$Po in the bulk of copper by using a commercial alpha-particle detector (Ultra-Lo 1800, XIA) \cite{210Po-source}.
The alpha detector has a good energy resolution (as explained in Sec.$\;$3.2) and a mechanism to reduce the background by waveform analysis, and thus its sensitivity is ${\sim10^{-4}\;\rm{\alpha/cm^2/hr}}$.
However, it has no position sensitivity.
A sample such as a micro pattern gas detector board 
does not have a
uniform radioactive contamination.
For example the impurities can be in a particular location due to the manufacturing process.
Therefore, a position-sensitive alpha detector is required in order to determine the site and perhaps the process associated with the materials contamination.

This paper is organized as follows.
The details of the alpha-particle detector, setup, low-$\alpha$ micro pixel chamber ($\mu$-PIC), gas circulation system, electronics, and trigger and data acquisition systems are described in ${\rm{Sec.\;2}}$.
The performance check that uses the alpha-particle source, a sample test, and background estimation are described in ${\rm{Sec.\;3}}$.
The remaining background of the detector and future prospects are discussed in ${\rm{Sec.\;4}}$.
Finally, 
main conclusions are presented in ${\rm{Sec.\;5}}$.

\section{Alpha-particle imaging detector based on gaseous micro-TPC}
\begin{figure}[tbh]
\includegraphics[width=5 cm, bb= 0 20 620 1020]{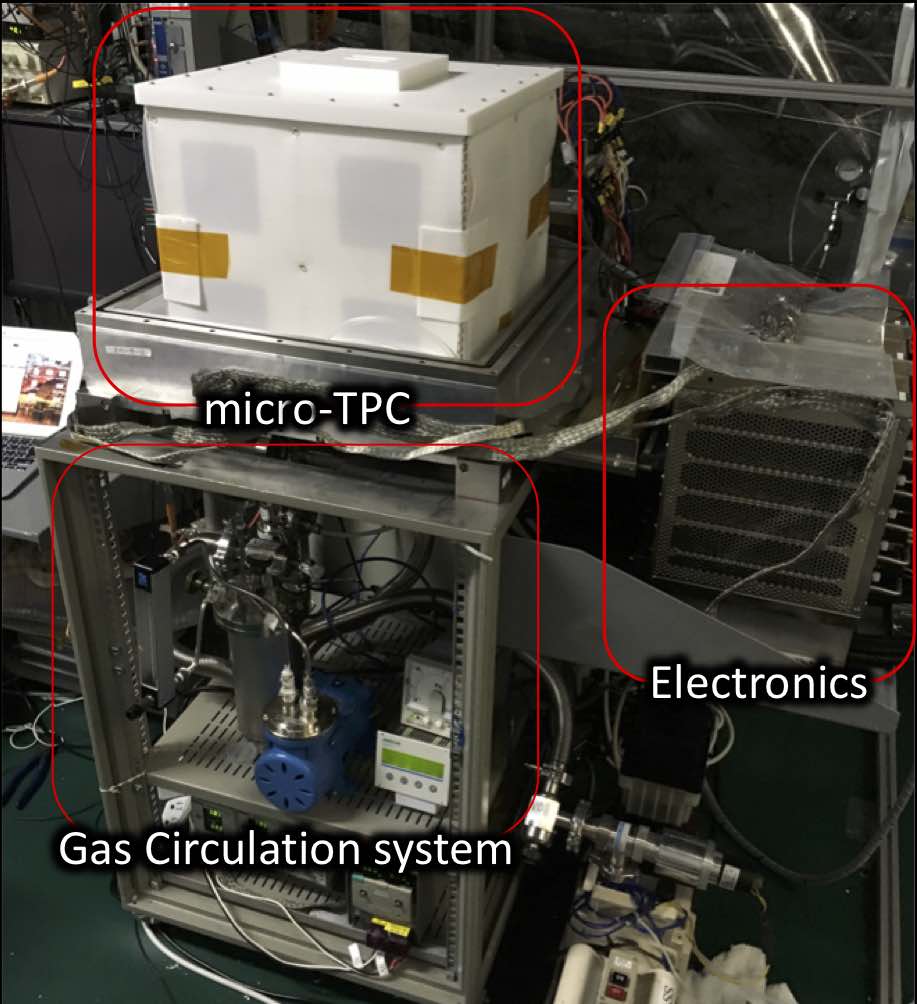}
\caption{
Photograph
of the experimental setup.
The detector system is composed of a micro-TPC, a gas circulation system, and electronics. The stainless-steel vessel is uncovered so that the outer view of the TPC field cage can be viewed.
}
\label{figure_detector_overview}
\end{figure}

A new alpha-particle detector was developed based on a gaseous micro-TPC upgraded from the NEWAGE-0.3a detector \cite{NEWAGE2010} which was used to search for dark matter from September, 2008 to January, 2013.
The detector consisted of the micro-TPC using a low-$\alpha$ $\mu$-PIC as readout, a gas circulation system, and electronics, as shown in Fig.\ref{figure_detector_overview}. The TPC was enclosed in a stainless-steel vessel for the gas seal during the measurement.

%
%

%
%
\subsection{\rm{Setup and configuration}}
\begin{figure}[tbh]
\includegraphics[width=5 cm, bb= 0 0 700 800]{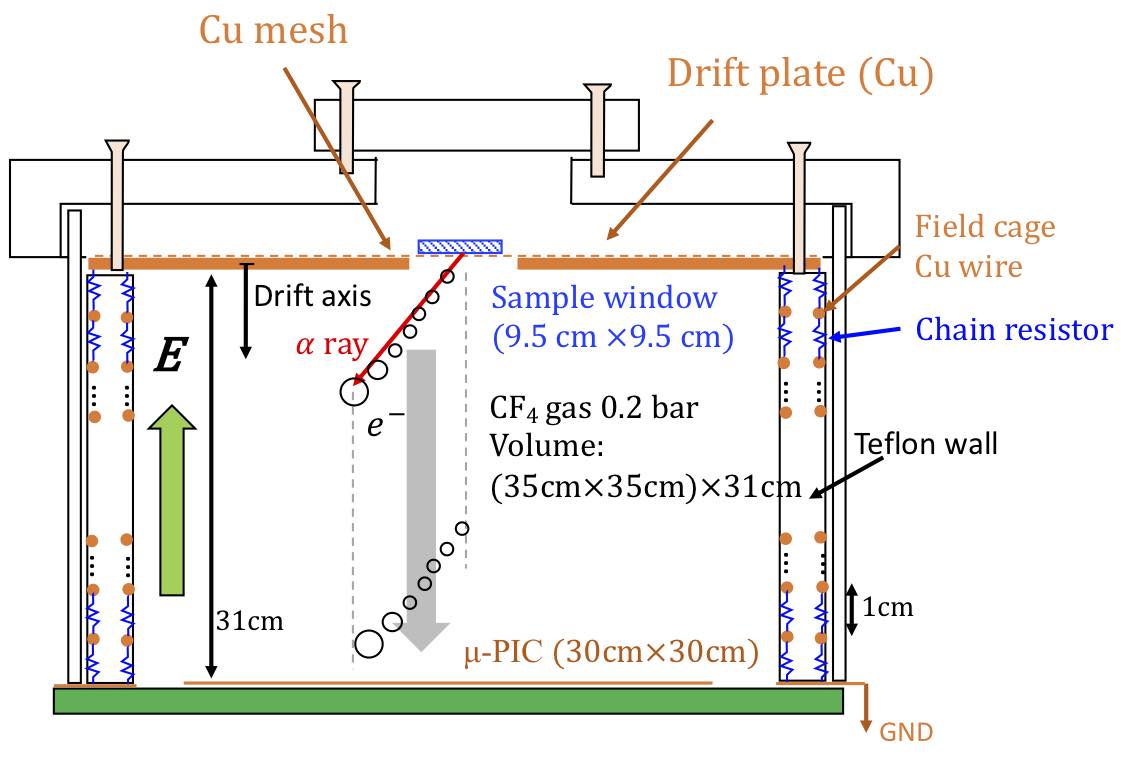}
\caption{
Schematic cross section of detector setup.
Sample window size is ${\rm 9.5\;cm\times9.5\;cm}$.
Electric field is formed by a drift plate biased at -$2.5\;{\rm kV}$ and copper wires with $1\;{\rm cm}$ pitch connecting with chain registers.
}
\label{TPC.figure}
\end{figure}

\begin{figure}[tbh]
\includegraphics[width=5 cm, bb= 0 0 750 900]{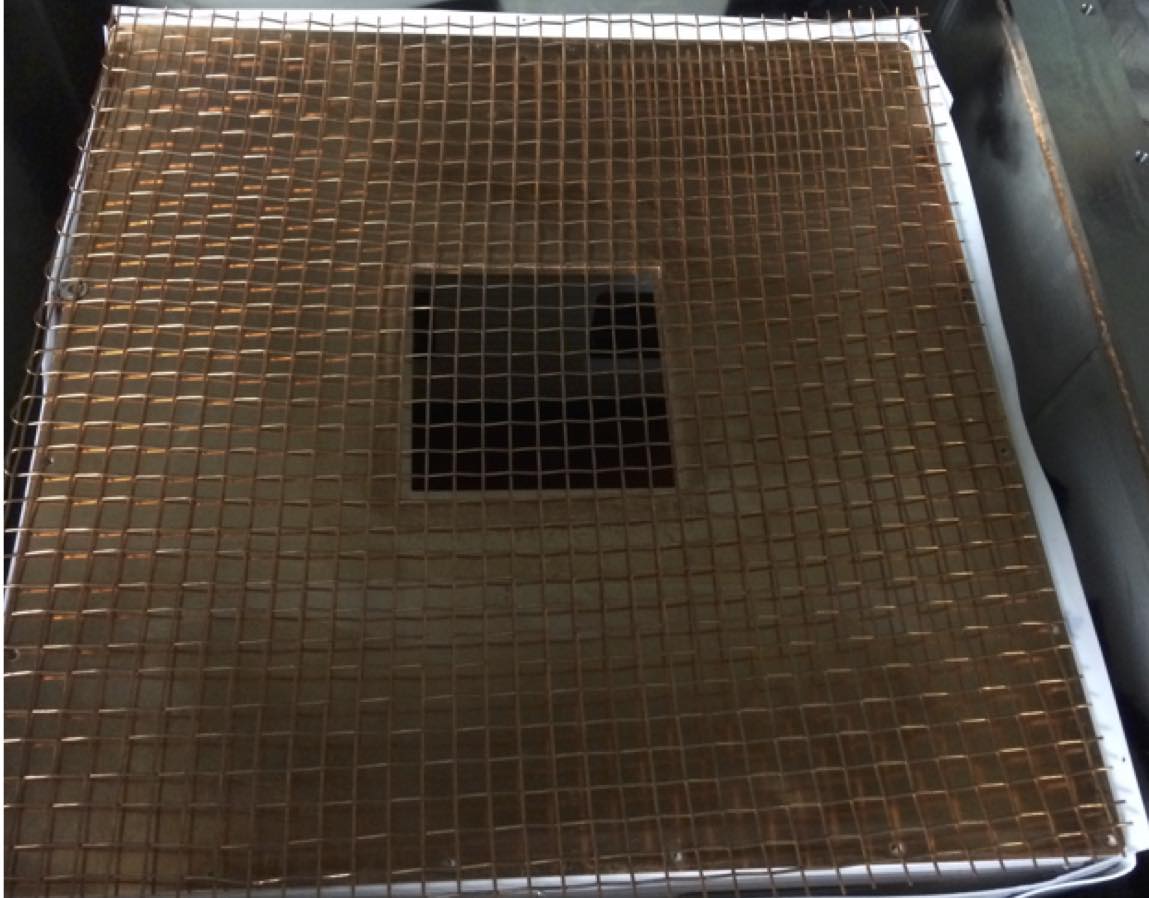}
\caption{Drift plate with a sample window (hole size is 
${\rm 9.5\;cm\times 9.5\;cm}$) and copper support mesh.
}
\label{drift.copper.mesh}
\end{figure}

Figure \ref{TPC.figure} shows a schematic view of the detector, where the gas volume is
${\rm (35\;cm\times 35\;cm)\times 31\;cm}$.
The detector was placed underground at the Kamioka facility in the Institute for Cosmic Ray Research, Japan.
An oxygen-free copper plate with a surface electro-polished to a roughness of ${0.4\;\rm{\mu m}}$ and a size of 
${\rm (35\;cm\times35\;cm)\times0.1\;cm}$ was used as the drift plate.
The drift plate had an opening with a size of ${\rm 9.5\;cm\times9.5\;cm}$ as a sample window.
A copper mesh made of 1-mm-$\phi$ wire in 1-cm pitch (aperture ratio of 0.81) was set on the drift plate to hold the sample at the window area, as shown in Fig.$\;$\ref{drift.copper.mesh}.
The electrons ionized by the alpha particles drift towards the $\mu$-PIC with a vertical upward-pointing electric field ${E}$.
$\rm{CF_4}$ gas (TOMOE SHOKAI Co.LTD, 5N grade: a purity of 99.999\% or more), which was also used in the NEWAGE-0.3a, was used
because of the low diffusion properties.
The pressure was set at ${0.2\;\rm{bar}}$ as a result of the optimization between the expected track length and the detector stability. 
The track length was expected to be longer, which improved the tracking performance when the gas pressures were low, while the discharge rate of the $\mu$-PIC increased.
The range of 5 MeV alpha particle is $\sim$8 cm in $0.2\;{\rm bar}$ $\rm{CF_4}$ gas, 
which would provide a reasonable detection efficiency considering the detector size.
The electric field in the drift volume, ${E=0.4\;\rm{kV/cm/bar}}$, was formed by supplying a negative voltage of ${2.5\;\rm{kV}}$ and placing field-shaping patterns with chain resistors every centimeter \cite{NEWAGE2007}.
The drift velocity was ${7.4\pm0.1\;\rm{cm/\mu s}}$.
The $\mu$-PIC anode was connected to ${+550\;\rm{V}}$.
The typical gas gain of $\mu$-PIC was $10^3$ at ${\sim500\;\rm{V}}$.

%
%

%
%

%
%
\subsection{\rm{Low-$\alpha$ $\mu$-PIC}}
The background study for the direction-sensitive dark matter search suggests that $\mu$-PIC has radioactive impurities of $^{238}\rm{U}$ and $^{232}\rm{Th}$ which emit alpha particles \cite{NEWAGE2015}.
A survey with a HPGe detector revealed that $\mu$-PIC's glass cloth was the main background source, and so the impurities were removed.
The polyimide with glass cloth in the $\mu$-PIC was replaced with a new material of polyimide and epoxy.
Details of the device with the new material, a low-$\alpha$ $\mu$-PIC, will be described in Ref \cite{LowAuPIC2018, Hashimoto.in.preparation}.

%
%
\subsection{\rm{Gas circulation system}}
A gas circulation system that uses activated charcoal pellets (Molsievon, X2M4/6M811) was developed for
the
suppression of radon background and a prevention of gain deterioration due to the outgassing.
A pump (EMP, MX-808ST-S) and a needle-type flow-meter (KOFLOC, PK-1250) were used to flow the gas at a rate of ${\sim500\;\rm{cm^3/min}}$.
The gas pressure was monitored to ensure the stable operation of the circulation system, operating within $\pm2\%$ for several weeks.

%
%
\subsection{\rm{Electronics and trigger and data acquisition systems}}
The electronics for the $\mu$-PIC readout consisted of amplifier-shaper discriminators \cite{ASD2004} for 768 anode and 768 cathode signals and a position-encoding module \cite{encorder2003} to reconstruct the hit pattern.
A data acquisition system consisted of a memory board to record tracks and a flash analog-to-digital converter (ADC) for the energy measurement.
The flash ADC with 100 MHz sampling recorded the sum signal of the cathode strips with a full time range of ${12\;\rm{\mu s}}$.
The anode sum signal issued the trigger.
The trigger
occurred when the electrons closest to the detection plane (indicated with the largest circle ($e^-$) in Fig.\;\ref{TPC.figure}) reach the $\mu$-PIC.
Since the main purpose of the detector is the alpha particle detection from the sample, the emission position of the alpha particle in the anode-cathode plane was determined at the position most distant from the $\mu$-PIC in the track (the smallest circle in Fig.$\;$\ref{TPC.figure}).

%

\section{Performance check}

%
%
\subsection{\rm{Alpha-particle source}}
A ${\rm 10\;cm\times10\;cm}$ 
copper plate with $^{210}$Pb accumulated on the surface was used as an alpha-particle source for the energy calibration and energy-resolution measurement \cite{210Po-source}.
The source emits alpha particles with an energy of ${5.3\;\rm{MeV}}$ as a decay of ${^{210}\rm{Po}}$.
The alpha-particle emission rate (hereinafter called the $\alpha$ rate) of the entire source plate was calibrated to be ${1.49\pm0.01\;\rm{\alpha\;s^{-1}}}$ for 4.8--${5.8\;\rm{MeV}}$ by using the Ultra-Lo 1800 \cite{210Po-source}.

%
\subsection{\rm{Energy calibration}}
\begin{figure}[tbh]
\includegraphics[width=4 cm, bb= 0 0 380 700]{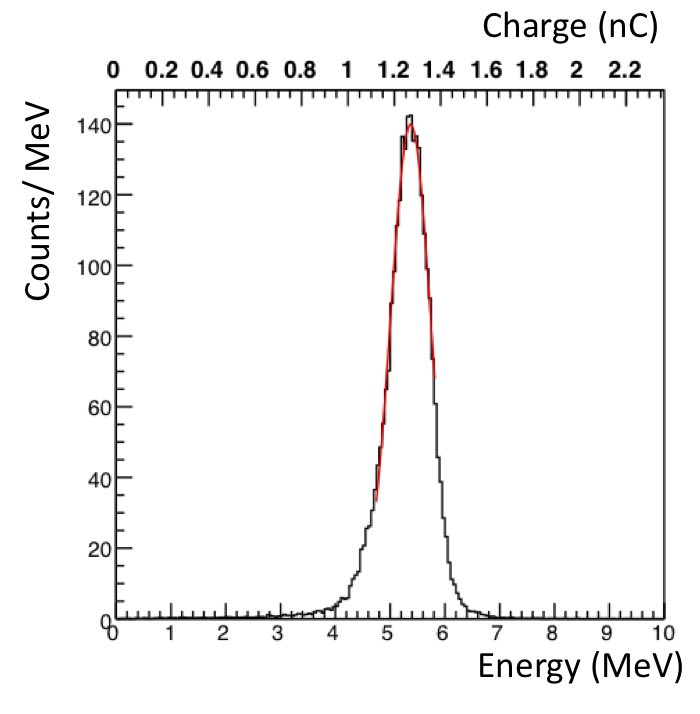}
\caption{Energy spectrum for alpha particles from $^{210}$Po (${5.3\;\rm{MeV}}$).
Red line is a fit result with a Gaussian.
}
\label{energy}
\end{figure}

An energy calibration was conducted with the alpha-particle source (5.3 MeV).
The event's energy was obtained by integrating the charge from the pulses registered by the 
flash
ADC. Thus spectra showed in this paper are presented in MeV.
Figure$\;$\ref{energy} shows a typical energy spectrum of the alpha-particle source. 
The energy resolution was estimated to be $6.7\%$ ($1\sigma$) for ${5.3\;\rm{MeV}}$, which is
not as good as
the Ultra-Lo 1800 resolution of 4.7\% ($1\sigma$) for 5.3 MeV. This deterioration was thought to be due to the gain variation of the $\mu$-PIC detection area.

%
%
\subsection{\rm{Event reconstruction}}
\begin{figure}[tbh]
\includegraphics[width=5.4 cm, bb= 0 0 580 800]{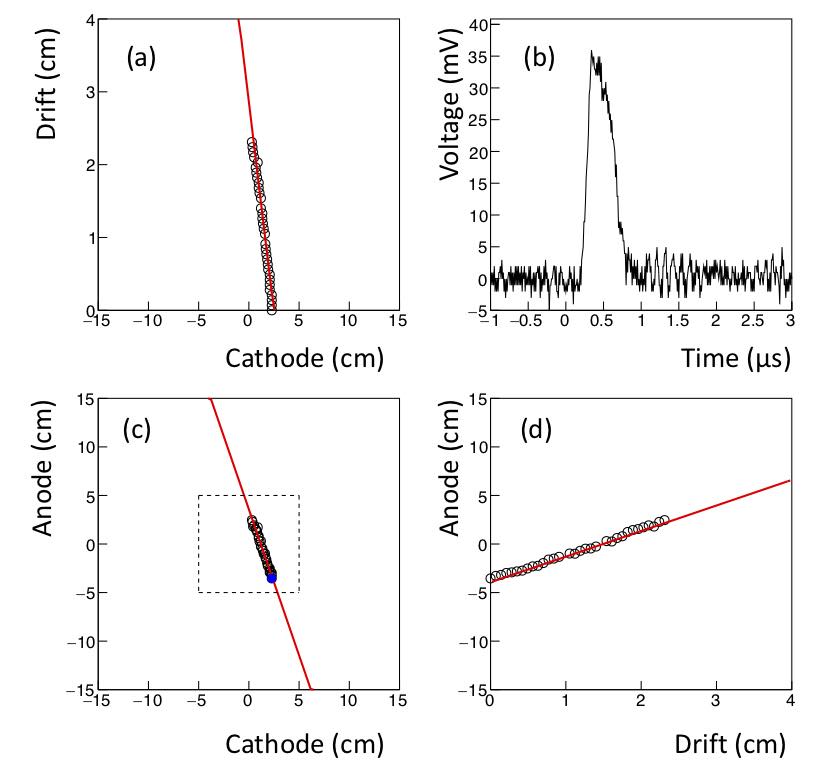}
\caption{
Event display of an alpha particle from 210Po. (a) cathode-drift projection, (b) flash ADC waveform (c) cathode-anode projection, and (d) anode-drift projection are displayed. The drift coordinate is set to zero base corresponding to the drift plate position for the top of the track.
}
\label{event.display}
\end{figure}

Figure \ref{event.display} shows a typical event display with the tracks and flash ADC waveform data for alpha-particle emission from $^{210}$Po.
The hit points were determined based on coincidence of anode and cathode detections.
Figure$\;$\ref{event.display} (c) shows the anode-cathode plane for the track.
The open circles correspond to hits registered in data.
The red solid line is a linear fit result.
The dashed line represents the edge of the sample window.
The solid blue point is the emission point of the alpha particle.
The scheme of the determination of the emission point, or the track sense, is explained in Sec.$\;$3.4.
Figure$\;$\ref{event.display} (a) and (d) show anode- and cathode-drift planes, respectively.
The drift coordinate is converted from the timing and is set to zero base, which corresponds to the drift-plate position.
Figure$\;$\ref{event.display} (b) shows a flash ADC waveform.

The track angles were determined on the anode-cathode, anode-drift, and cathode-drift planes.
These angles were determined with a common fitting algorithm.
First, the weighted means of the hit points $(x_{\rm{w}},y_{\rm{w}})$ were defined as
\begin{eqnarray}
  \left(
    \begin{array}{c}
      x_{\rm{w}}  \\
      y_{\rm{w}}  \\
    \end{array}
  \right)=\frac{1}{n}\sum^{n}_{j=0}{
  \left(
    \begin{array}{c}
      x_j  \\
      y_j  \\
    \end{array}
  \right)},
\end{eqnarray}
where $x_j$ and $y_j$ are the measured hit points and $n$ is the number of points.
Next, the track was shifted and rotated through the angle $\theta$ as follows
\begin{eqnarray}
  \left(
    \begin{array}{c}
      x'_j  \\
      y'_j  \\
    \end{array}
  \right)=
    \left(
    \begin{array}{cc}
      \cos\theta & -\sin\theta \\
      \sin\theta & \cos\theta \\
    \end{array}
  \right)
  \left(
    \begin{array}{c}
      x_j - x_{\rm{w}} \\
      y_j - y_{\rm{w}} \\
    \end{array}
  \right).
\end{eqnarray}

%
%
Here $x'_j$ and $y'_j$ are the points after the 
shift, the rotation angle
$\theta$ were determined to minimize the quantity $f$, which is defined as
\begin{eqnarray}
f(\theta) = \sum{{y'}_{j}^{2}},
\label{eq.1}
\end{eqnarray}
where this formula means a sum of the square of the distance between the rotated point and the $x$ axis.
This method has the advantage to determine the angle with no infinity pole at 
${\theta=90^{\circ}}$
(i.e. parallel to cathode strip (fitting in the anode-cathode plane) or drift axis (fitting in the anode-drift and cathode-drift plane)).

%
%

%
%
\subsection{\rm{Track-sense determination}}
Backgrounds in low radioactivity alpha-particle detectors are in general alpha particles from the radon (radon-$\alpha$) and materials of construction used in the detector (detector-$\alpha$).
The radon-$\alpha$'s are expected to be distributed uniformly in the gas volume with isotropic directions. The detector-$\alpha$'s are expected to have position and direction distributions specific to their sources.
One of the main sources of the detector-$\alpha$'s 
is the $\mu$-PIC so the directions of {$\alpha$'s} coming from this component are mostly upward-oriented.
Since the direction of alpha particles from the sample are downward, these detector-$\alpha$'s and half of the radon-$\alpha$'s can be rejected by the cut of upward-direction events.

The deposit energy per unit path length, $dE/dx$ of an alpha particle with an initial energy over a few MeV, has a peak before stopping (Bragg peak).
The number of electrons ionized by the alpha particle in the gas is proportional to $dE/dx$, and $dE/dx$ along the track profile is projected onto the time evolution in the signal due to the mechanism of the TPC.
This time profile was recorded as the waveform and thus the track sense (i.e., whether the track was upward or downward) can be determined from the waveform.

\begin{figure}[tbh]
\includegraphics[width=5.5 cm, bb= 0 0 400 900]{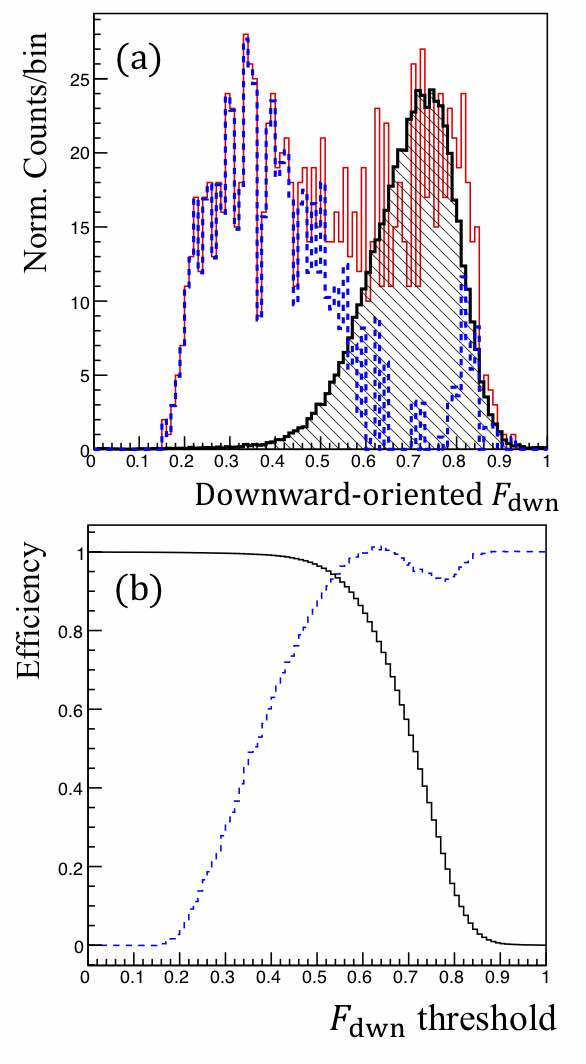}
\caption{(a) Downward-oriented distribution for source-$\alpha$ (black shade), radon-$\alpha$ (red solid), and a histogram made by subtracting the radon-$\alpha$ spectrum from the source-$\alpha$ one(blue dashed) 
(b) Detection efficiency for downward-(black solid) and rejection efficiency for upward-oriented (blue dashed) events as a function of $F_{\rm{dwn}}$ threshold.
}
\label{upwardlike}
\end{figure}

A parameter to determine the track sense is
\begin{eqnarray}
F_{\rm{dwn}}=S_2/(S_1+S_2),
\end{eqnarray}
where $S_1$ and $S_2$ are the time-integrated waveform before and after the peak. They are defined as
\begin{eqnarray}
&&S_1=\int^{t_p}_{t_0}{v(t) dt},\\
&&S_2=\int^{t_1}_{t_p}{v(t) dt}.
\end{eqnarray}
Here, ${t_0=0\;\rm{\mu s}}$, ${t_1=1.5\;\rm{\mu s}}$, and $t_p$ are the start, stop, and peak time, respectively, for the waveform shown in Fig.$\;$\ref{event.display}\;(b).
The $t_p$ is determined as a time when the voltage is highest in the region between $t_0$ and $t_1$.
Figure$\;$\ref{upwardlike} (a) shows typical $F_{\rm{dwn}}$ distribution with the alpha-particle source, where most of the events are expected to be downward-oriented.
The $F_{\rm{dwn}}$ values of the downward events are distributed around 0.7, as shown by the black-shaded histograms.
Conversely, radon-$\alpha$'s have an isotropic direction, 
i.e., $F_{\rm{dwn}}$ has two components of upward- and downward-oriented,
as shown by the red solid histogram, where the radon-$\alpha$ are background events in the sample test data, as explained later.
The scale of the source-$\alpha$ was normalized to the radon-$\alpha$ peak of downward for clarity.
Figure 6 (b) shows the efficiency related on $F_{\rm{dwn}}$ threshold for downward-(black solid) and upward-oriented (blue dashed).
The selection efficiency of ${F_{\rm{dwn}}>0.5}$ was estimated to be $0.964\pm0.004$ in the source-$\alpha$ spectrum while the radon background was reduced to half.
The blue dashed histogram is a spectrum that subtracted the normalized source-$\alpha$ from the radon-$\alpha$. The cut efficiency of the upward-oriented events (${F_{\rm{dwn}}\leq0.5}$) was estimated to be ${0.85\pm0.04}$.
The energy dependence of $F_{\rm{dwn}}$ will be explained in Sec. 3.6.

%
%
\subsection{\rm{Distribution of emission position}}
\begin{figure}[tbh]
\includegraphics[width=5.5 cm, bb= 0 0 400 1000]{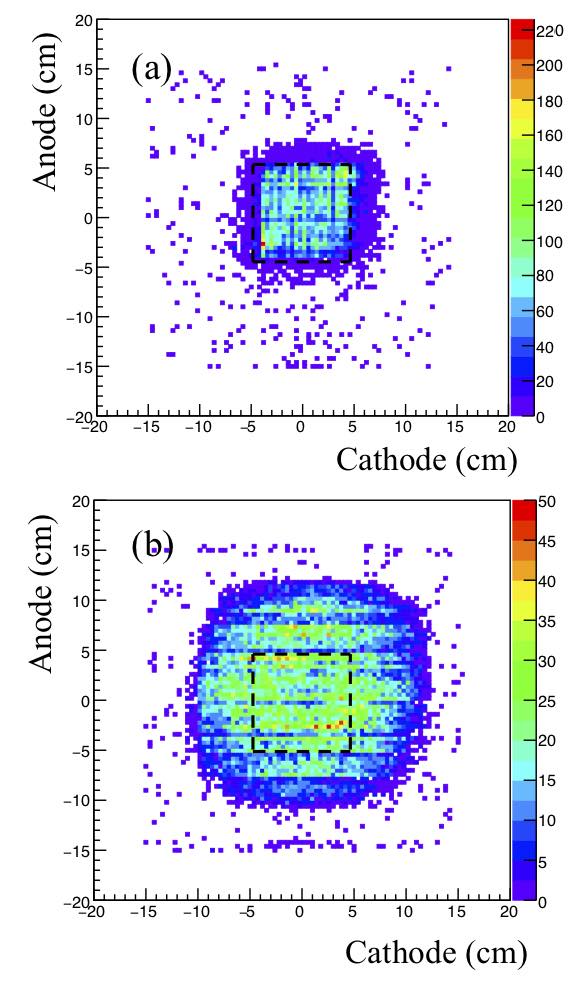}
\caption{Anode--cathode projection distributions of (a) top and (b) bottom of tracks for alpha particles emitted from the source. The dashed line is the edge of the sample window.}
\label{alpha.map}
\end{figure}

Since alpha particles are mainly emitted from the source, the top points of the alpha-particle tracks trace the shape of the radioactivity on the sample.
Figures$\;$\ref{alpha.map} (a) and \ref{alpha.map} (b) show the anode--cathode projection distribution of the top and bottom of the alpha-particle tracks, respectively,
where the top and bottom are defined as the zero and maximum drift coordinate, respectively, as shown in Fig.$\;$\ref{event.display} (a) and \ref{event.display} (d).
The dashed line represents the edge of the drift-plate sample window.
Comparing Fig.$\;$\ref{alpha.map} (a) with Fig.$\;$\ref{alpha.map} (b) clearly reveals the shape of the radioactivity.

\begin{figure}[tbh]
\includegraphics[width=5.5 cm, bb= 0 0 800 1100]{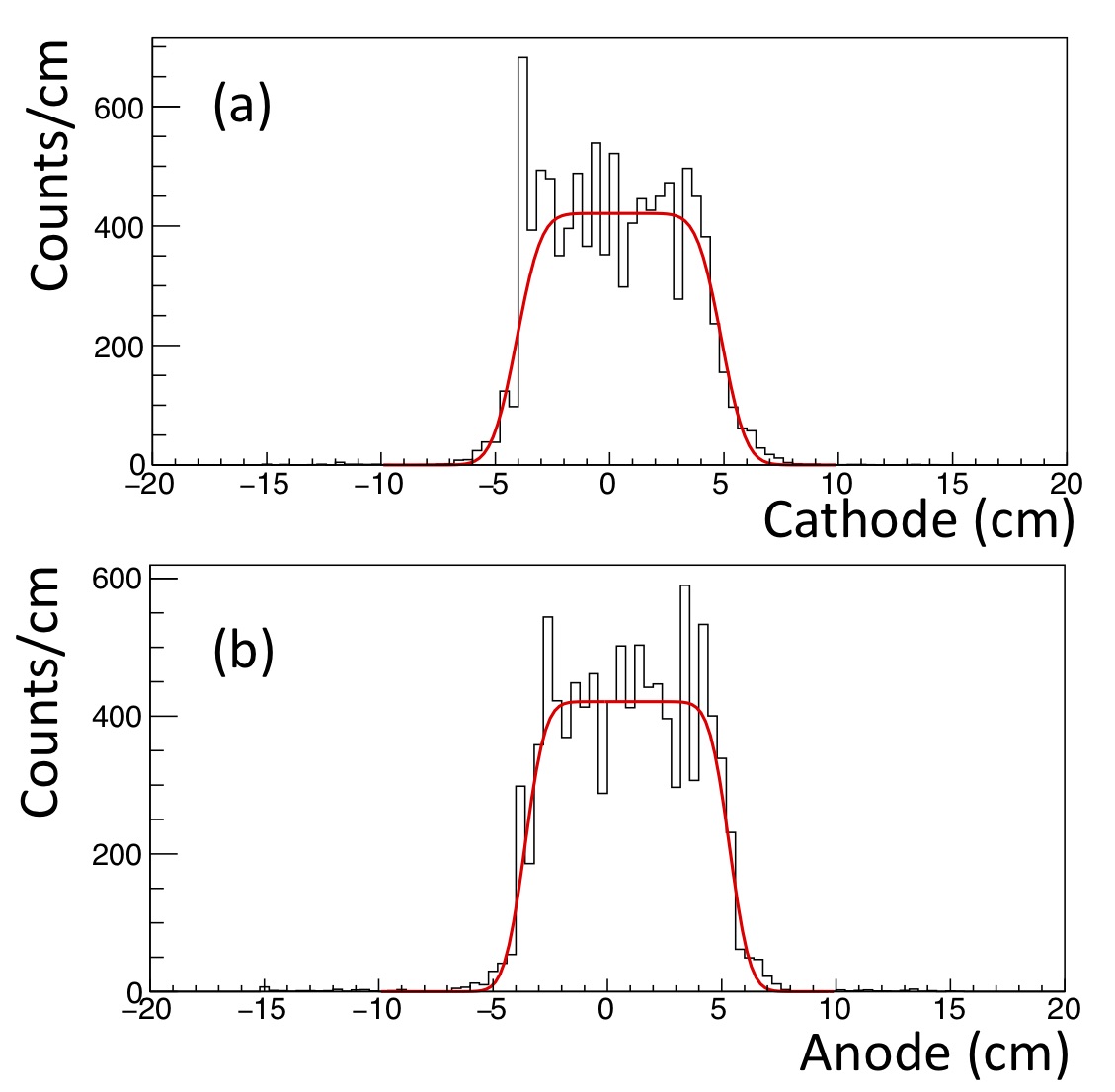}
\caption{
Alpha-particle emission position projected to cathode (a) and anode (b). Red lines represent fitting with error functions.
}
\label{reso.est}
\end{figure}

The position resolution was evaluated along the four dashed lines in Fig.$\;$\ref{alpha.map} (a).
The number of events was projected onto the axis perpendicular to the lines and was fit with error functions as shown in Fig.$\;$\ref{reso.est}.
Figure $\;$\ref{reso.est} (a) and (b) represent the alpha-particle emission position projection to cathode and anode, respectively. The red lines are the fitting based on the error functions.
As a result, the position resolution was determined to be ${0.68\pm0.14\;\rm{cm}}\;(\sigma)$, where the error is a standard deviation in the four positions.

%
%
\subsection{\rm{Detection and selection efficiency}}
To select good events for alpha particles from the sample, we use the following criteria: (C1) selection for events with good fitting tracks, (C2) cut for the upward-oriented events, and (C3) selection for events with emission points in the sample region.

For criterion C1, the
best
fit to track events was selected as $f_{\rm{min}}(\theta)/(n-1)<0.02\;\rm{cm^2}$.
It
was
determined as the best $\theta$ to minimize $f(\theta)/(n-1)$ at each plane, for both
tracking of electrons
and $\alpha$-ray. The electron track tends to be scattered, so $f_{\rm min}(\theta)/(n-1)$
for electrons
is bigger than that of $\alpha$-ray. Therefore, the upper limit of $f_{\rm min}(\theta)/(n-1)$ serves to suppress electron-track events.

Criterion C2 rejects the upward-oriented tracks with ${>3.5\;\rm{MeV}}$ and ${F_{\rm{dwn}}\leq0.5}$ because the determination efficiency depends on the energy.
The upward- and downward-oriented tracks can be determined with ${95\%}$ or more certainly at over ${3.5\;\rm{MeV}}$.
Note that this cut was applied for the events ${>3.5\;\rm{MeV}}$, because the radon background, which was assumed to be the dominant background source, created the peak around ${6\;\rm{MeV}}$ and the contribution to the energy range below ${3.5\;\rm{MeV}}$ was limited.

For criterion C3, the source-$\alpha$ was selected within a 
region of $\pm$8 cm in both the anode and cathode. The cut condition was decided to cover both tails of the distribution
(or $>4\sigma$)
in Fig.$\;$\ref{reso.est} (a) and (b).
The rate of radon-$\alpha$ in the selected region was around two orders of magnitude lower than the source-$\alpha$ rate,
and considered negligible.

The selection efficiency for C1, C2, and C3 containing the detection efficiency was calculated to be ${(2.17\pm0.29)\times10^{-1}\;\rm{counts/\alpha}}$ (the ratio of the count rate to the $\alpha$ rate of the source), where the error represents the systematic error of C1 to C3 selections and 
uncertainty of the source radioactivity is considered negligible.

\subsection{\rm{Sample test and background estimate}}

\subsubsection{\rm{Setup}}
\begin{figure}[tbh]
\includegraphics[width=6 cm, bb= 0 0 470 470]{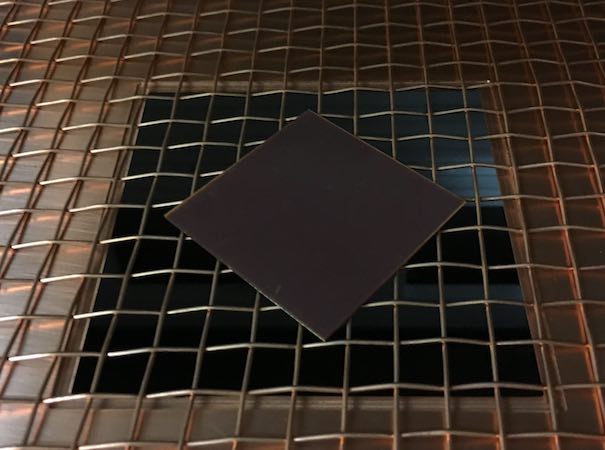}
\caption{Setup for a 
${\rm 5\;cm\times5\;cm}$
piece of the standard $\mu$-PIC as sample.}
\label{sample.uPIC}
\end{figure}

A ${\rm 5\;cm\times5\;cm}$
piece of the standard $\mu$-PIC whose $\alpha$ rate was known to be $0.28\pm0.12\;\rm{\alpha/cm^2/hr}$ in previous work \cite{LowAuPIC2018} served as a sample and was inspected by using the detector.
A photograph of the sample position over the setup mesh is
shown in Fig.$\;$\ref{sample.uPIC}.
The measurement live time
was 75.85 hr.

\subsubsection{\rm{Background in sample region}}

The $\alpha$ rate of the sample was estimated by subtracting the background rate.
Considered background was mainly the radon-$\alpha$.
The detector measured both the $\alpha$ rates in the region of the sample and around the sample (outer region).
The background rate could be determined from the $\alpha$ rate in the outer region.
Recall, 
the upward and downward radon-$\alpha$ rates are same.
The sample-$\alpha$ has mainly downward-oriented.
Thus, 
the background rate could be estimated
by the upward rate in the sample region and independently cross-checked by the upward rate in the outer region.

\begin{figure}[tbh]
\includegraphics[width=6 cm, bb= 0 0 440 550]{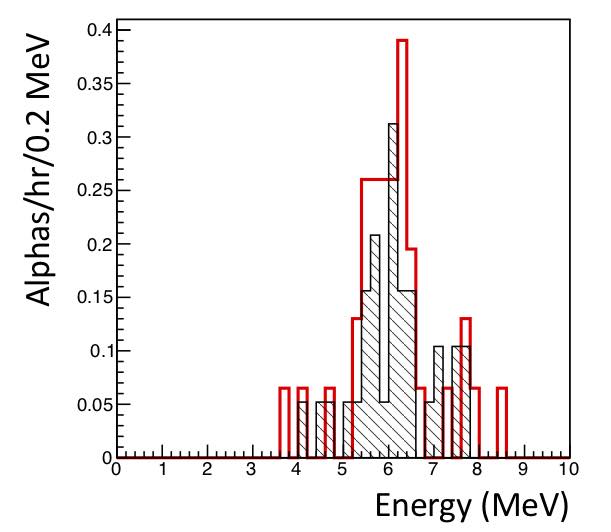}
\caption{
Upward-oriented
alpha-particle energy spectra inside (red) and outside (black shade) the sample region.
The dashed line is the threshold of 2 MeV.
}
\label{sample.radon}
\end{figure}

We checked the upward-oriented (${F_{\rm{dwn}}\leq0.5}$) ${\alpha\;\rm{rate}}$ in both regions because the alpha particles from a sample are typically emitted downward.
Measured energy spectra are shown in Fig.$\;$\ref{sample.radon}.
The red- and black-shaded histograms show the energy spectra inside and outside the sample region, respectively.
These spectra are scaled by the selection efficiency.
Both peaks are around ${6\;\rm{MeV}}$ and $\alpha$ rates are
${(2.16^{+0.54}_{-0.35})\times10^{-2}}$ (inside) and 
${(1.54^{+0.64}_{-0.40})\times10^{-2}\;\rm{\alpha/cm^2/hr}}$ (outside).
Therefore, the background condition inside the sample region 
is compatible at less than 1$\sigma$
with the background condition outside the sample region.
The alpha-particle energy spectrum is interpreted as the radon peaks at ${5.5\;\rm{MeV}}$ ($^{222}$Rn), ${6.0\;\rm{MeV}}$ ($^{218}$Po), and ${7.7\:\rm{MeV}}$ ($^{214}$Po).

The downward-oriented (${F_{\rm{dwn}}>0.5}$) $\alpha$ rate outside the sample is ${(1.58^{+0.29}_{-0.26})\times10^{-2}\;\rm{\alpha/cm^2/hr}}$, as shown in the black-shaded spectrum of Fig.$\;$\ref{sample.ene}.
In this work, the background rate was improved by one order of magnitude in comparison with that of our previous work \cite{LowAuPIC2018}.
The background reduction is attributed to the track-sense determination to reject upward-oriented alpha (for ${>3.5\;\rm{MeV}}$) and the replacement of the low-$\alpha$ $\mu$-PIC (for ${\leq3.5\;\rm{MeV}}$).
In the energy region between 2.0 and ${4.0\;\rm{MeV}}$, where most radon background is suppressed, the background rate is $(9.6^{+7.9}_{-5.6}){\times10^{-4}\;\alpha/\rm{cm^2/hr}}$.

\subsubsection{\rm{$\alpha$ rate of sample}}

\begin{figure}[h]
\includegraphics[width=6 cm, bb= 0 0 450 600]{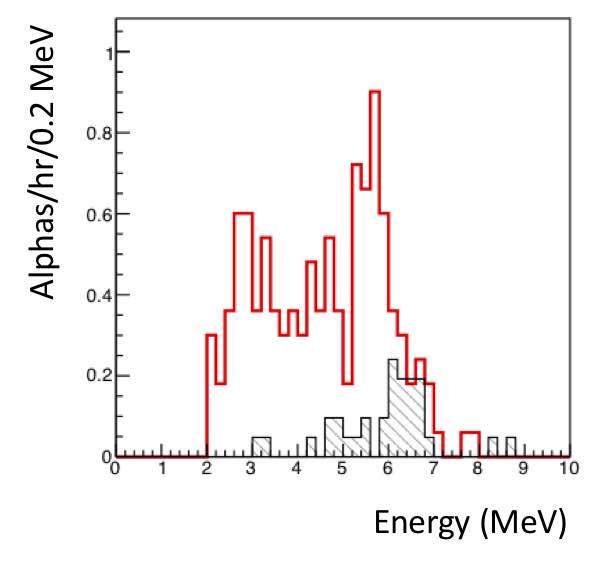}
\caption{
Downward-oriented alpha-particle energy spectra in sample region (red) and background region (black shade).
The dashed line is the threshold of 2 MeV.
}
\label{sample.ene}
\end{figure}

\begin{figure}[h]
\includegraphics[width=6 cm, bb= 0 0 450 520]{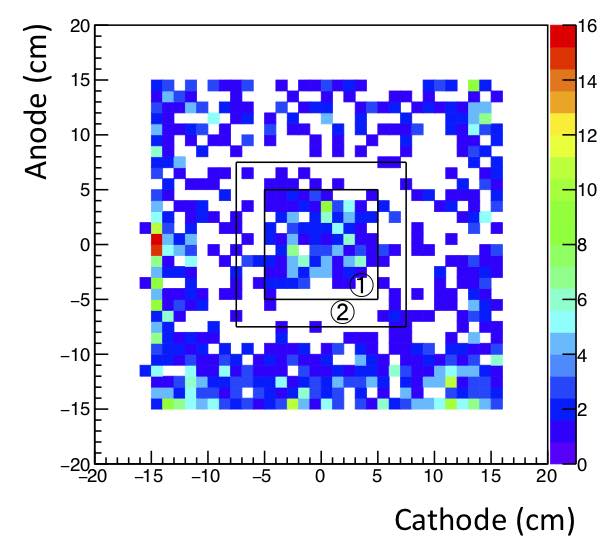}
\caption{
Distribution of the top of downward-oriented alpha-particle track. The regions $\textcircled{\scriptsize 1}$ and $\textcircled{\scriptsize 2}$ are the sample and background regions, respectively.
}
\label{sample.map}
\end{figure}

Figure$\;$\ref{sample.map} shows the distribution of the top of the tracks for the sample, where the candidates are selected by the criteria C1 and C2.
The regions $\textcircled{\scriptsize 1}$ and $\textcircled{\scriptsize 2}$ are defined as sample and background regions, respectively. The sample region corresponds to the sample window.
The sample region is the inside of $\pm5\;\rm{cm}$ of anode and cathode.
The background region is the outside of the sample region and the inside of $\pm7.5\;\rm{cm}$ of anode and cathode.
The systematic uncertainty due to the setting of the background region is estimated by changing the outer bound by $\pm$ 0.5cm to be $\sim0.5\%$.
Figure$\;$\ref{sample.ene} shows the energy spectra of downward-oriented alpha particles in the sample (red) and the background region (black shaded).
The $\alpha$ rate of the sample was calculated to be
${(3.57^{+0.35}_{-0.33})\times10^{-1}\;\rm{\alpha/cm^2/hr}}$ ($>2.0\;\rm{MeV}$) 
by subtracting the background rate.

Here, the impurity of $^{232}$Th and $^{238}$U is estimated by comparing with a prediction of $\alpha$ rate spectrum in the simulation, where it mentions that the isotope in the material is assumed as only $^{232}$Th or $^{238}$U because of the continuous $\alpha$ rate spectrum. In the fit region between 2 and 10 MeV, the impurity of $^{232}$Th or $^{238}$U is estimated to be $6.0\pm1.4$ or $3.0\pm0.7$ ppm, respectively.
The impurities of $^{232}$Th and $^{238}$U are measured to be $5.84\pm0.03$ and $2.31\pm0.02$ ppm, respectively, by using the HPGe detector
with the measuring time of 308 hr.
Although the error is huge because of the continuous energy spectrum, it is consistent with the prediction of prior measurement.
In this sample test, it was demonstrated to observe the background alphas at the same time.

\begin{table*}[htbp]
  \centering
  \begin{tabular}{lcc}\hline
     & This work & HPGe detector \\ \hline\hline
    Sample volume (${\rm cm}$)& $(5\times5)\times0.098$ & $(5\times5)\times2.47$ \\
    Sample weight (g) & 6.8 & 169.5\\
    Measureing time (hr) & 75.85 & 308\\
    Net $\alpha$ rate ($\alpha/{\rm cm^2/hr})$) & $(3.57^{+0.35}_{-0.33})\times10^{-1}$ & --- \\
    $^{232}$Th impurities (ppm)  & $6.0\pm1.4$ &$5.84\pm0.03$\\
    $^{238}$U impurities (ppm) & $3.0\pm0.7$ & $2.31\pm0.02$\\\hline
  \end{tabular}
  \caption{Comparison of Screening result with this work and HPGe detector.}
  \label{result}
\end{table*}

%
%
\section{Discussion}
We begin by discussing the sensitivity for the energy between 2 and ${9\;\rm{MeV}}$ based on long-term measurements.
In this energy range, the background is dominated by the radon-$\alpha$'s with $\sim(1.58^{+0.29}_{-0.26})\times10^{-2}\;\rm{\alpha/cm^2/hr}$.
The statistical error $(\sigma)$ is expected to scale with the inverse of the square root of the measurement time $(t)$ given as ${\sigma \propto 1/\sqrt{t}}$.
In this work, the live time was only three days, and the statistical error was ${\sigma\sim3\times10^{-3}\rm{\alpha/cm^2/hr}}$.
With a measurement time of one month, the error of sample-$\alpha$'s was estimated to be ${\sigma\sim1\times10^{-3}\;\rm{\alpha/cm^2/hr}}$.
The sensitivity for a sample with a radioactivity much lower than the background rate is practically determined by the statistics of the background when the background can be subtracted. The expected statistical errors of both the background and sample are ${1\times10^{-3}\;\alpha/\rm{cm^2/hr}}$ with one month of measurement time. The statistical error of the subtracted event rate, or the detection sensitivity of the sample, is therefore expected to be a few ${\times10^{-3}\;\alpha/\rm{cm^2/hr}}$.

\begin{figure}[tbh]
\includegraphics[width=5 cm, bb= 0 0 600 700]{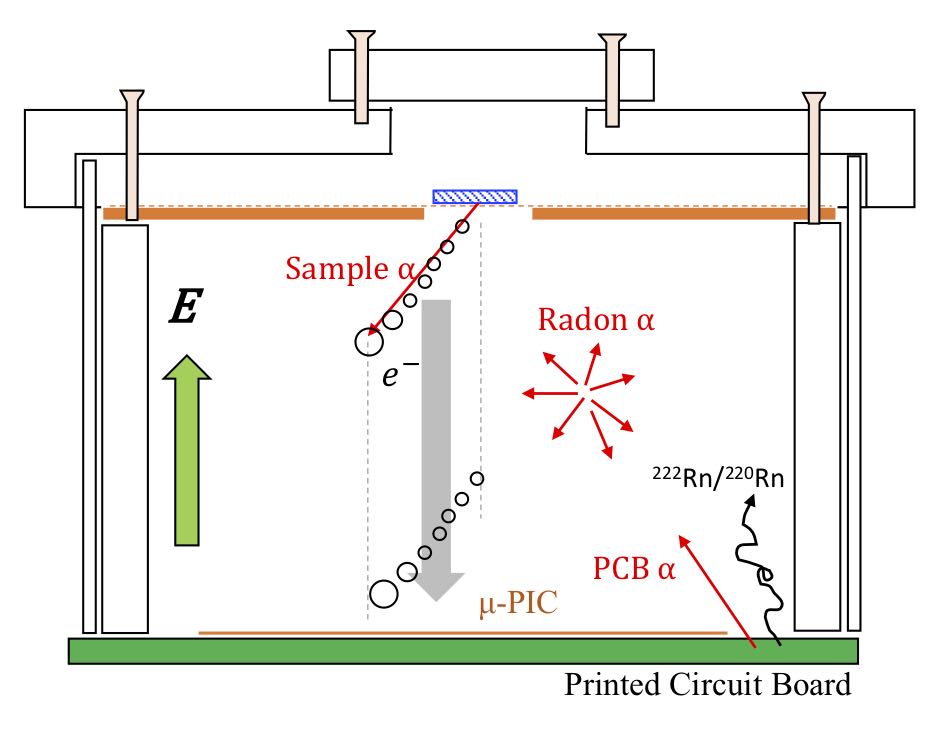}
\caption{Schematic cross section of background alpha particles in detector setup.}
\label{BG.TPC.figure}
\end{figure}

The edges region (anode ${\sim\pm15\;\rm{cm}}$ or cathode ${\sim\pm15\;\rm{cm}}$) has a high rate of background, as shown in Fig.$\;$\ref{sample.map}.
These events have an energy and path-length dependence similar to that of the alpha particles.
The alpha particles were mainly oriented upward and were emitted from outside the detection area, limited by the $\mu$-PIC.
As an impurity candidate, a piece of the printed circuit board (PCB) was inspected and the $\alpha$ rate was ${(1.16\pm0.06)\times10^{-1}\;\rm{\alpha/cm^2/hr}}$.
Although the alpha-particle events could be rejected by the fiducial region cut, these impurities could be the radon sources (see Fig.$\;$\ref{BG.TPC.figure}).
Therefore, as a next improvement, a material with less radiative impurities should be used for the PCB.

%

The goal for detector sensitivity is less than ${10^{-4}\;\rm{\alpha/cm^2/hr}}$. We can potentially reduce the background rate by using the cooled charcoal to suppress radon gas and using a material with less impurities. Insulators such as polytetrafluoroethylene, polyimide, and polyetheretherketone, are in general low radioactive if we can use them without extra materials with relatively high radioactive like reinforcing glass-cloth.
A recent study reported that a cooled charcoal could suppress the radon by 99\% in the argon gas \cite{RADIOISOTOPES2010}.
A recent NEWAGE detector suppresses the radon to 1/50 by using cooled charcoal \cite{NEWAGE2015}.
With these improvements, the detector would achieve to the goal of performance.

%
%
\section{Conclusion}
We developed a new alpha-particle imaging detector based on the gaseous micro-TPC.
The measured energy resolution is ${6.7\%}$ ($\sigma$) for ${5.3\;\rm{MeV}}$ alpha particles.
The measured position resolution is ${0.68\pm0.14\;\rm{cm}}$.
Based on a waveform analysis, the downward-oriented events' selection efficiency is ${0.964\pm0.004}$ and the cut efficiency of the upward-oriented events is ${0.85\pm0.04}$ at ${> 3.5 \;\rm{MeV}}$.
Also, a piece of the standard $\mu$-PIC was measured as a sample, and the result is consistent 
with the one obtained by a measurement done with a HPGe detector.
A measurement of the alpha particles from a sample and background was also established at the same time.
A background rate near the radon-$\alpha$ (${(1.58^{+0.51}_{-0.42})\times10^{-2}\;\rm{\alpha/cm^2/hr}}$) was achieved.

%
%
\section*{Acknowledgments}
This work was supported by a Grant-in-Aid for Scientific Research on Innovative Areas, 26104004 and 26104008, from the Japan Society for the Promotion of Science in Japan.
This work was supported by the joint research program of the Institute for Cosmic Ray Research (ICRR), the University of Tokyo.
We thank Dr. Y. Nakano of the ICRR, University of Tokyo, Japan for providing us with a helium-gas leak detector.

%
%

\bibliography{mybibfile}
\end{document}